\documentclass[conference]{IEEEtran}
\usepackage{cite}

\ifCLASSINFOpdf
\else
\fi
\usepackage{cleveref}

\usepackage{url}

\usepackage{graphicx}

\usepackage{subcaption}
\usepackage{tabu}

\usepackage{amsthm}
\usepackage{float}
\usepackage{algorithm}

\usepackage{algorithmic}
\usepackage[usenames,dvipsnames]{xcolor}

\usepackage{booktabs}

\begin{document}
%
\title{Gwardar: Towards Protecting a Software-Defined Network from Malicious Network Operating Systems}

\author{\IEEEauthorblockN{Arash Shaghaghi\IEEEauthorrefmark{1}\IEEEauthorrefmark{2}, Salil S. Kanhere\IEEEauthorrefmark{1}, Mohamed Ali Kaafar\IEEEauthorrefmark{2}\IEEEauthorrefmark{3} and Sanjay Jha\IEEEauthorrefmark{1}}
\IEEEauthorblockA{\IEEEauthorrefmark{1}The University of New South Wales (UNSW Sydney), Australia\\
\IEEEauthorrefmark{2}CSIRO Data61, Australia \\
\IEEEauthorrefmark{3}Macquarie University, Australia \\
\{a.shaghaghi, salil.kanhere, sanjay.jha\}@unsw.edu.au\\
dali.kaafar@data61.csiro.au}
}


%


\maketitle

\begin{abstract}
A Software-Defined Network (SDN) controller (aka. Network Operating System or NOS) is regarded as the brain of the network and is the single most critical element responsible to manage an SDN.  Complimentary to existing solutions that aim to protect a NOS, we propose an intrusion protection system designed to protect an SDN against a controller that has been successfully compromised. Gwardar maintains a virtual replica of the data plane by intercepting the OpenFlow messages exchanged between the control and data plane. By observing the long-term flow of the packets, Gwardar learns the normal set of trajectories in the data plane for distinct packet headers. Upon detecting an unexpected packet trajectory, it starts by verifying the data plane forwarding devices by comparing the actual packet trajectories with the expected ones computed over the virtual replica. If the anomalous trajectories match the NOS instructions, Gwardar inspects the NOS itself. For this, it submits policies matching the normal set of trajectories and verifies whether the controller submits matching flow rules to the data plane and whether the network view provided to the application plane reflects the changes. Our evaluation results prove the practicality of Gwardar with a high detection accuracy in a reasonable time-frame.

\end{abstract}

\begin{IEEEkeywords} 
Software-Defined Network;
SDN Security;\ifpdf
Controller Security
\end{IEEEkeywords}

\IEEEpeerreviewmaketitle

\section{Introduction \& Background}


In the era of cyber-war and cyber-terrorism, attackers are targeting the very core of today's network infrastructure~\cite{nsa2,nsa}. Attacks against the network infrastructure have three key properties, which make them attractive for attackers: 1) are hard to detect, 2) take a long time to fix, and 3) once successful, may grant the most unrestricted type of access to information \cite{shaghaghisurvey}.


Software-Defined Network (SDN) has emerged as a fundamental re-design of traditional networks with increased adoption in the past few years. Numerous proposals have emerged that leverage its capabilities to either improve existing network-based services or create new ones~\cite{ali2015survey}. These proposals assume a secure and trusted SDN platform, which has motivated an active area of research aiming to do so. \par

Securing SDN's layered architecture is not straightforward, mainly due to the incompatibility of existing solutions \cite{dhawan2015sphinx} and the added number of threat vectors. As discussed in~\cite{kreutz2015software}, the most dangerous SDN specific threat vectors are a) compromised network controllers and b) malicious SDN applications. In fact, a successful attack at SDN control plane (or, `the brain of the network') could potentially grant an attacker the capability to control the entire network and target both the service provider and its users. \par

Existing solutions proposed to secure the control plane of an SDN mainly protect a controller when interacting with other layers. Solutions such as~\cite{shin2014rosemary, scott2014operationcheckpoint, banse2015secure} aim to secure an SDN controller against malicious applications installed at the application plane by applying advanced access control restriction mechanisms.  Another line of research is focused on the secure composition of policies submitted by different SDN applications (e.g. \cite{rezvani2016anomaly, jin2015covisor}). The SDN Appstore ecosystem and the different types of SDN controllers (e.g. centralized vs. decentralized, different architectures and designs, implementation languages and ongoing updates) pose an on-going challenge for these solutions. \par

On the other hand, an increasing number of studies aim to secure the southbound API and data plane layer of SDNs. For instance, an attacker can manipulate the controller's view of the network by tampering with the OpenFlow messages exchanged between a controller and the network forwarding devices. Alternatively, a malicious forwarding device can execute a Denial of Service (DoS) attack against its controller and take down an SDN. We provide an extensive survey in \cite{shaghaghisurvey}. \par

Here, we focus on the scenario where the controller has been maliciously compromised by an attacker either by compromising the SDN controller software (also known as the Network Operating System or NOS) or by violating the access privileges. The latter may be due to the presence of a malicious administrator (i.e. an insider) or malicious applications installed on a controller. In fact, a NOS is hosted on a commodity server and may be subject to software hacks as any other application.  For instance, an SDN rootkit \cite{ropke2015sdn} grants an attacker the capability to adversely re-program a network while concealing its attempts. To the best of our knowledge, only a few solutions have been proposed against such a powerful adversary and most SDN security solutions assume a secure and trusted NOS. Tatang et al. have proposed SDN-GUARD \cite{tatang2017sdn}, which aims to detect malicious network programming attempts by an SDN rootkit. However, the threat model used in this work is specific to SDN rootkits and the proposed solution assumes that the NOS is not compromised. \par




We propose Gwardar, an Intrusion Protection System (IPS) designed to protect an SDN against compromised SDN controllers. Gwardar retrieves the packet trajectories from the data plane and creates normal models for packet trajectories traversing the network forwarding devices. Gwardar also maintains a virtual replica of the network by intercepting the OpenFlow messages exchanged between the control and data plane. This virtual replica is used to verify the normal models created for the packet trajectories. Whenever derivations are detected, Gwardar first performs a trajectory-based inspection of the forwarding devices by employing the attack detection algorithms of WedgeTail \cite{shaghaghi2017wedgetail}. For this, it compares the suspicious packet trajectories `with the expected ones computed over the virtual replica that it maintains to detect and locate possible malicious forwarding devices. Thereafter, if the anomaly matches the rules specified by the control plane, it inspects the control plane. For this, Gwardar submits flow rules matching the normal set of trajectories to the controller with a high priority and evaluates whether: a) the controller submits the flow rules correctly to the data plane, and b) the controller updates the global network view available to applications after these changes. Gwardar detects a compromised NOS when any of the conditions above are invalid. Gwardar may be programmed by its administrator on how to respond to threats. By default, however, it retrieves rules from the most valid virtual replica copy it maintains and applies the valid flow rules to remove the malicious trajectories. In extreme cases, Gwardar may be programmed to take over the network until the NOS has been fixed.


Hereon, we discuss the threat model that Gwardar is degined against in \S\ref{threatmodel}, present an overview of its architecture in \S\ref{overview}, and evaluations results in \S\ref{implementEvaluate}. We conclude the paper in \S\ref{conclusion}.






\section{Threat Model \& Assumptions} \label{threatmodel}

We consider one of the least explored but strongest adversaries against an SDN, where an attacker has managed to compromise the NOS either partially or completely. No restrictions are assumed on how the attacker has compromised the controller and we focus on detecting and preventing the malicious actions. The adversary's goal is to install malicious flow rules to drop, replay, misroute and modify the packets in a random or selective manner affecting all or part of the traffic being routed through the data plane. Moreover, the NOS is assumed to be capable of concealing these actions from the upper layers. Hence, the malicious behaviours cannot be detected just by querying the NOS for global network view. \par
We assume that the SDN has been trustworthy and secure long enough time for Gwardar to learn the normal behaviour of the network (i.e. has not been compromised since day one). Lastly, to the best of our knowledge, we are the first work that assumes an attacker may have compromised more than one layer of an SDN. In fact, when detecting threats Gwardar inspects both the network data plane and control plane. Lastly, we assume that the application plane and the communications through the northbound and southbound APIs are secure.

\section{Architecture of Gwardar} \label{overview}
\begin{figure*}[!htb]
\setlength\belowcaptionskip{-15pt}
  \centering	
  \includegraphics[width=5.7in, scale=0.50, height=2.35in]{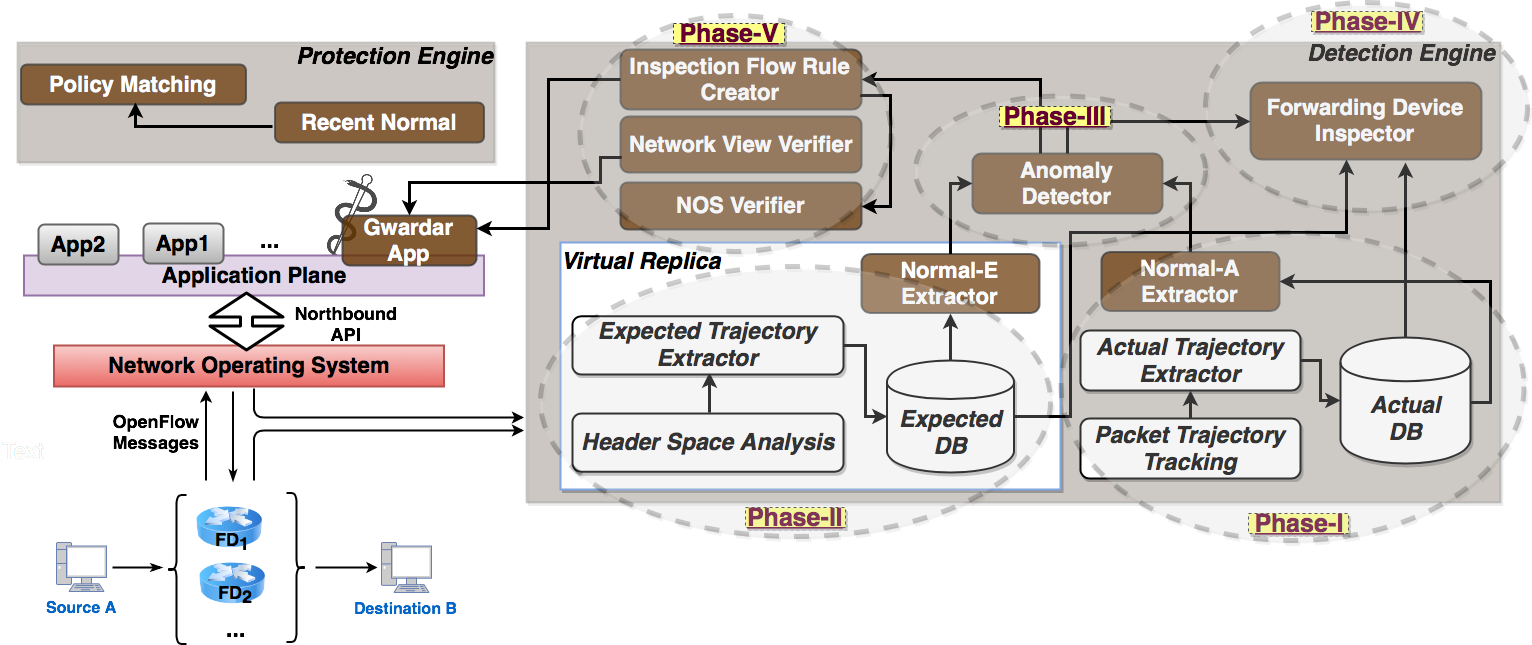}
  \caption{Gwardar Architecture}
  \label{architecture}
\end{figure*}

As shown in Figure \ref{architecture}, Gwardar is composed of two main components: Detection Engine and Protection Engine. In a nutshell, the former aims to detect threats while the latter is designed to rectify issues and if not possible, take over as a temporary acting NOS. \par

\subsection{Detection Engine} The detection engine has five main phases. \textit{Phase-I:} Retrieving the actual packet trajectories, identifying the scanning regions and computing the normal packet trajectory models for these regions. \textit{Phase-II:} Intercepting the OpenFlow messages exchanged between the control and data plane to maintain a virtual replica of the network over a Mininet network. It also involves computing the expected packet trajectories over the virtual replica, which match the normal packet trajectory models retrieved in \textit{Phase-I}. \textit{Phase-III:} The anomaly detection phase takes as input the output of prior phases and verifies the normal packet trajectory models to detect anomalies. \textit{Phase-IV:} Upon detection of anomalies, Gwardar starts by inspecting the data plane forwarding devices. For this, it employs the attack detection algorithms used in \cite{shaghaghi2017wedgetail}. This process involves retrieving trajectories that a packet may take against each of the other forwarding devices and comparing them with the actual trajectories retrieved from the network. Gwardar's Expected DB and Actual DB are used to provide these trajectories. \textit{Phase-V:} If the anomalous trajectories match the flow rules submitted by the control plane, Gwardar moves to the NOS inspection phase. This process involves: a) creating flow rules matching the normal model trajectory, b) intercepting and verifying the associated OpenFlow FlowMod messages submitted by controller to data plane, and c) querying the network view and verifying whether it reflects the changes submitted. \par

\subsection{Protection Engine} \par
Gwardar's protection engine reacts to threats detected by the detection engine in order to apply temporary fixes. In fact, if an attacker manages to take over an SDN controller it becomes a very resourceful adversary and it becomes rather difficult to restore the network to normal conditions without the intervention of a trusted administrator. Hence, the following two mechanisms are envisioned for Gwardar's protection engine: a) immediate protective forwarding policies to protect the critical services and resources, and b) taking over all network configurations using the most recent trusted configurations. For the former, Gwardar's administrator defines the policies. The policy specification and implementation mechanisms are similar to the `Response Engine' of WedgeTail \cite{shaghaghi2017wedgetail}. In the second case, Gwardar retrieves the forwarding device configurations from the most recent and trusted network snapshot it maintains and applies them by sending FlowMod messages to the network. No further updates are applied to the network hereon.

\subsection{Gwardar's Main Components} \label{comparch}

Gwardar builds on several core functional features we proposed in our earlier work WedgeTail \cite{shaghaghi2017wedgetail}. These include creating a virtual network replica by intercepting OpenFlow messages exchanged between the control and data plane (`Virtual Replica'), retrieving expected packet trajectories for specific packet headers (`Expected Trajectory Extractor'), tracking packets over the network and creating packet trajectories (`Actual Trajectory Extractor'). These components have been used as the building blocks of Gwardar and provide input to its different components. In the following, we succinctly present these components. \par

\textbf{Normal Extractors} The Normal Extractor components are responsible to process the trajectory databases (i.e. `Actual DB' and `Expected DB'). `Normal-A Extractor' has two main functions: I) identifying the scanning regions, and II) computing the normal model for packet trajectories for each scanning region. For the former, Gwardar keeps track of trajectories for all packets on all ports over time and identifies the most commonly involved forwarding devices by looking at the denser regions. For this purpose, as also in \cite{shaghaghi2017wedgetail} we use Unsupervised Trajectory Sampling technique \cite{pelekis2010unsupervised}. Algorithm 1 represents how the normal model is calculated. In a nutshell, the normal model is time dependent and composed of the normal model of all different trajectories in each scanning region. The process involves breaking down each packet trajectory in a scanning region into smaller time-dependent trajectories and retrieving the forwarding devices involved in these trajectories. These are then stored as the normal forwarding devices for a packet trajectory. Breaking down trajectories during this process makes calculating the normal model efficient. Moreover, it helps reducing the false positives by minimizing the impact of networking configurations such as load balancing, which might result into temporary trajectories. \par
Once the normal model is created, whenever there is a Forwarding Device (FD) at time $t+1$ that is not available in the normal set, the Anomaly Detector component is triggered.

\begin{algorithm}
\setlength\belowcaptionskip{-15pt}
 \caption{Normal Extractor for a trajectory}
 \begin{algorithmic}
 \renewcommand{\algorithmicrequire}{\textbf{Input:}}
 \REQUIRE $T_i$ $\in$ $SR_i$, Time $t$
\STATE List $Normal(T_i)$
  \FOR {all $FD(x)$ $\in$ $T_i$.ForwardingDevices()}
   \STATE List L = \{all FDs at least 2 hops away from $FD(x)$\}
    \FOR {all y = L.ForwardingDevices()}
    \STATE Packet $Pck$; 
    \STATE Time-dependent Trajectory $T_{j,t}$; 
    \STATE List Normal;
    \STATE $Pck$.Source() = $FD(x)$;
    \STATE $Pck$.Destination() = $FD(y)$;
    \STATE $Pck$ = Find-Packet($Pck$.Source, $Pck$.Destination);
        \STATE $T_{j,t}$ = Actual Trajectory ($Pck$);
    \STATE $Normal(T_i)$ =+ $T_{j,t}$.ForwardingDevices();
    \ENDFOR
  \ENDFOR
  
\end{algorithmic} 
\end{algorithm}
\setlength{\textfloatsep}{0pt}
\textbf{Anomaly Detector}
The anomaly detector takes as input the forwarding device detected as anomalous. It then requests the `Normal-E Extractor' component to to retrieve the expected packet trajectories using `Header Space Analysis' module for the packet leading to anomaly. It then inspects all the forwarding devices in the packet trajectory to detect the root cause. For this Gwardar uses the attack detection algorithms proposed in our earlier work \cite{shaghaghi2017wedgetail}, which would enable Gwardar to locate the potentially malicious forwarding device(s) and detect their exact malicious action including packet drop, misroute, and etc. \par
If the anomaly detector does not detect a forwarding device as the cause of the unexpected trajectory and this anomaly re-occurs, the NOS inspection phase is started.

\textbf{Inspection Flow Rule Creator \& NOS Verifier} The former submits a set of flow rules fixing the anomalous packet trajectory to the controller. This rule is then submitted to the NOS through the Gwardar SDN App with a high priority (i.e. it will not be replaced in the policy composition process). The NOS Verifier component listens for OpenFlow FlowMod messages submitted to the data plane and verifies whether these are correct. Note that the custom flow rules submitted by the controller are removed from the network once intercepted to ensure the data plane status is stable and not affected by Gwardar's checks. \par

\textbf{Network View Verifier} SDN Gwardar App queries for the network view from the NOS using the Northbound API and using the `Network View Verifier' module evaluates whether the network view reflects the updates submitted. The most recent flow rules matching the network conditions before inspection are resubmitted to the network to restore the network view after the inspection phase.

\section{Implementation \& Evaluation} \label{implementEvaluate}
Compared to solutions such as WedgeTail \cite{shaghaghi2017wedgetail} and SPHINX \cite{dhawan2015sphinx}, Gwardar is not envisioned to be installed as an application for SDN controllers. Instead, it is designed as a trusted third-party application with full access to an SDN platform responsible to verify it. In order to work, Gwardar does not require tampering with either the control or data plane and can be easily imported into different platforms. We implemented Gwardar mainly in Java and over simulated settings. We used Floodlight as the network controller and evaluated Gwardar over two simulation network settings, which were different in terms of the number of forwarding devices, network subnets, network rules and trajectories. Table 1 shows a comparison of the different simulation settings. \par
\textbf{Network Topologies, Flow Rules and Network Traffic:} Zib54 and Sprint setup network topologies were extracted from \cite{orlowski2010sndlib} and \cite{orlowski2010sndlib} respectively. To add flow-entries, we created an interface for a subset of prefix found in a full BGP table of Route Views \cite{routeview} and spread them randomly and uniformly to each router as `local prefixes'. We then computed forwarding tables using shortest path routing. Mausezahn \cite{Mausezahn} and custom scripts were used to add benign traffic to the simulated networks. As in \cite{dhawan2015sphinx, shaghaghi2017wedgetail}, our scripts imported real-world network traces from \cite{3,4} to drive traffic into Mininet.

\begin{table}[]
\centering
\setlength\belowcaptionskip{-1pt}
\begin{tabular}{ccc}
\hline
\textbf{Number of}         & \textbf{Zib54} & \textbf{Sprint} \\ \hline
\textbf{Forwarding Device} & 54         & 316        \\
\textbf{Subnet}            & 750        & 48751      \\
\textbf{Rules}             & 8752       & 7492131    \\
\textbf{Trajectory}        & 11723      & 252169     \\ \hline
\end{tabular}

\caption{Overview of simulated networks}
\end{table}

\subsection{Attack Scenarios \& Implantation}
We analyzed Gwardar's detection success over 250 different attack instances. These were randomly implanted using custom developed script matching the following six different scenarios: \par
\textbf{Scenario-I} Compromised NOS submitted malicious rules to one forwarding device, \textbf{Scenario-II} Compromised NOS submitted malicious rules to multiple forwarding devices, \textbf{Scenario-III} Compromised NOS submitted malicious rules to one forwarding devices and manipulated the network view to conceal actions, \textbf{Scenario-IV} same as Scenario-III but for multiple forwarding devices, \textbf{Scenario-VI} One or more malicious forwarding devices while the NOS was secure and trustworthy. The malicious actions of forwarding devices included packet drop, replay, misroute, and modification.

\subsection{Performance Analysis}
We measure Gwardar's performance with respect to successful detection of attacks and applying protections. We also report on the impact of packet losses in Gwardar's detection success as well as resource utilization and the user-perceived latency. \par

\textbf{Detection Accuracy and Performance} We report that Gwardar has been able to detect all the implanted attack scenarios over the simulated networks. We compute attack detection time as the absolute time taken by Gwardar to detect the anomaly and send a signal to the protection engine. This time does not include the time take to create the normal model as this is dependent on the network size and status. In our simulations for Sprint setup, as shown in Figure 2, the false positive rate reached a stable level after about 2 hours and a half. After this time the number of false positive reached less than 5\%. In those cases, Gwardar performs a full inspection of the data plane forwarding devices. Note that in our evaluations, we configured Gwardar to resume learning mode after detecting major network reconfigurations. \par

The attack distribution was as follows: 100 attacks over Zib54 setup and 150 attacks over Sprint setup. The average detection time over Zib54 Setup is about 235 seconds with a standard deviation of 9 seconds. For Sprint setup, the average detection time is 540 second with a standard deviation of 75 seconds. Hence, the detection time scales well with respect to the network size. In essence, it is satisfactory to a network administrator to automatically detect attacks against the network forwarding devices and NOS in about 10 minutes (after Gwardar's training time). Figure 3 illustrates a comparison of detection time for 50 attacks in Zib54 and Sprint setups.
\begin{figure}
\setlength\belowcaptionskip{-12pt}
    \centering
    \includegraphics[width=0.65\linewidth]{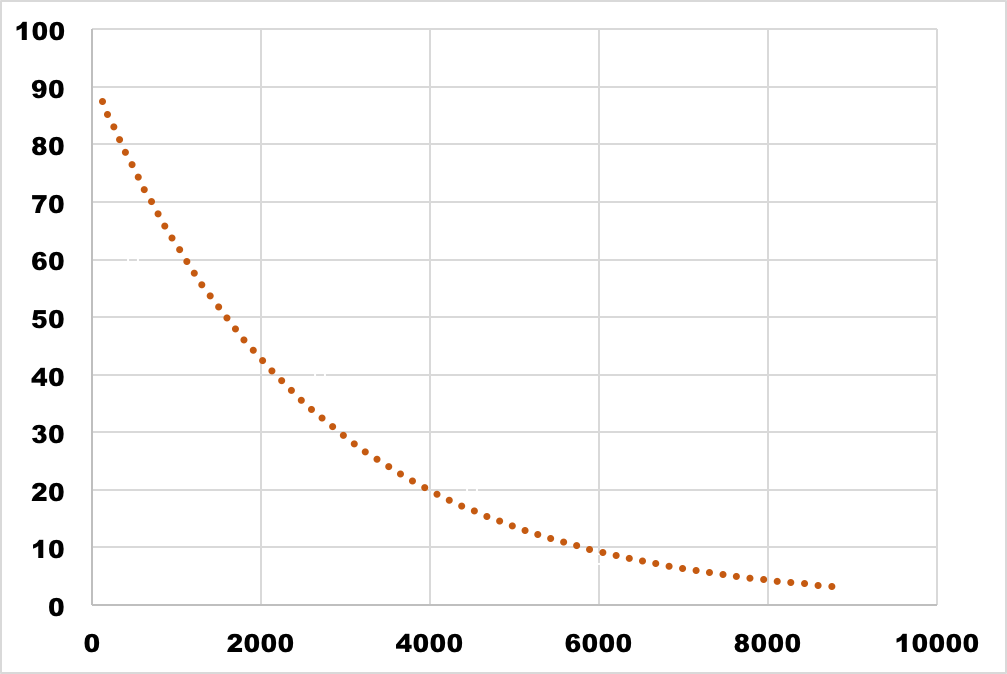}
    \caption{False positive adjustment in Sprint setup after about 3 hours of passively intercepting network traffic.}
    \label{fig:falsepositive}
\end{figure}

\begin{figure}
    \centering
    \includegraphics[width=0.65\linewidth]{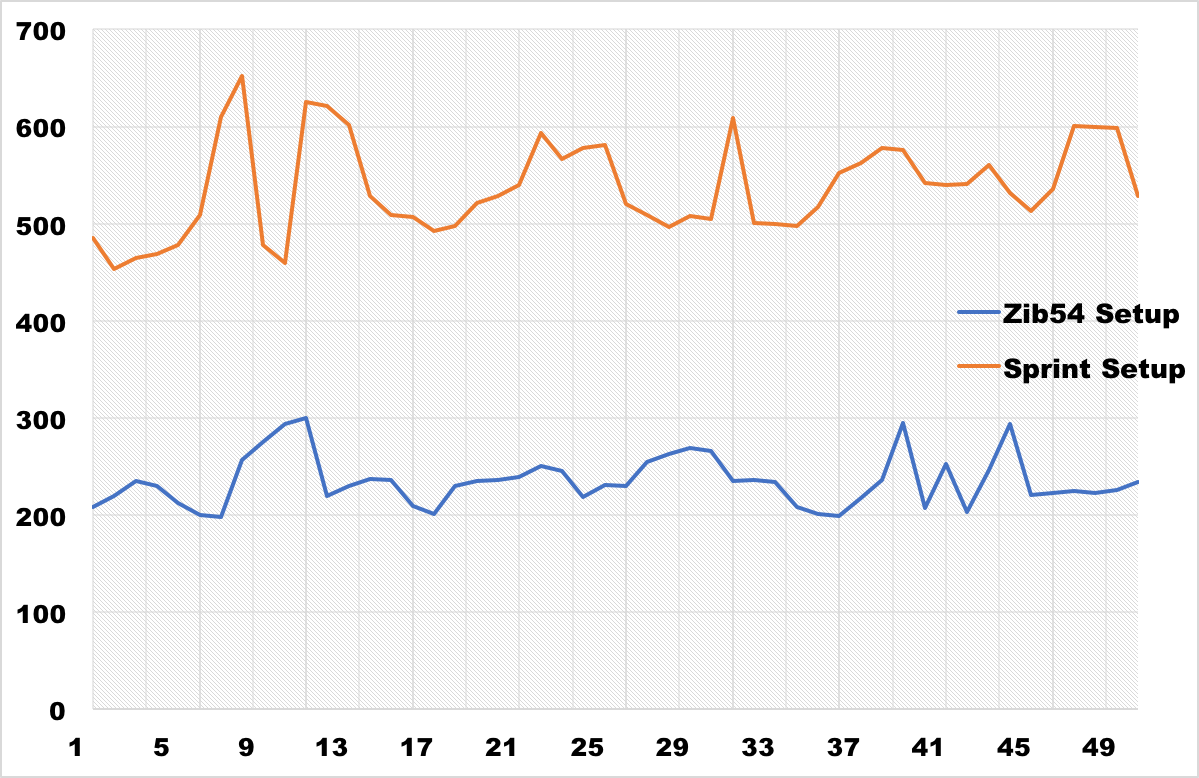}
    \caption{Comparison of Gwardar's performance when detecting threats over Zib54 and Sprint setups for 50 implanted attacks.}
    \label{fig:falsepositive}
\end{figure}


\textbf{Protection Performance} Gwardar's protection engine successfully applied the policies specified when threats were detected. If no policy was defined, the most recent replica of the network was retrieved and matching flow rules were applied. One of the issues we faced was that when applying flow rules from past to one or a set of forwarding devices, the network routing configurations became unstable leading to black and grey holes. Hence, it is recommended Gwardar to apply full restoration of the network or defining custom policies to avoid such issues during this process. A full restoration of network flow rules over Sprint setup took on average about 10 seconds.

\textbf{Impact of Packet Losses}
Gwardar's trajectory based mechanism is similar to \cite{shaghaghi2017wedgetail}. We address this challenge by employing \cite{mizrak2008detecting}, which allows n detecting packet drops or gray hole attacks in networks by exploiting the correlation between packet delays and packet losses due to congestion.

\textbf{Resource Utilization}
We hosted the simulated networks on a machine equipped with Intel Core i5, 2.66 GHz quad-core CPU and 16 GB of RAM. Gwardar was hosted on a machine with Intel Core i7, 2.66 GHz quad-core CPU and 8 GB of RAM. During the detection phase, the average CPU usage reaches a maximum of 20\% while the memory consumption reaches 25\%. 

\textbf{User Perceived Latency}
As a non-real time IDS, Gwardar does not have any performance implications during its detection phase. It is not possible to generalize its performance when preventing threats given that this is case specific. 

\section{Related Work}
As mentioned in \S1, we are not aware of any existing solution aiming to detect a compromised NOS. Given that Gwardar employs certain functions of our earlier work WedgeTail (see \S\ref{comparch}), we briefly highlight their key differences. \textit{First,} completely different scopes and threat models. WedgeTail aims to detect a compromised forwarding device while Gwardar aims to detect a compromised malicious NOS. \textit{Second,} WedgeTail scans the whole network and each and every forwarding device after retrieving the scanning regions. Instead, Gwardar analyzes the trajectory database and creates normal models of the network and inspects the forwarding devices as the first step upon detecting anomalies. \textit{Third,} completely different response capabilities. WedgeTail submits fixing policies as an application to NOS while Gwardar submits them directly to the network and is also capable of taking over the NOS altogether. 

\section{Conclusion} \label{conclusion}
A compromised NOS in an SDN grants an attacker limitless number of attack vectors allowing him/her to target the service provider and users. Complimentary to existing solutions aiming to secure the NOS platforms, we proposed Gwardar as an IPS designed to detect a compromised NOS based on the network data plane trajectories and independent of underlying software and hardware. Our solution is capable of detecting a compromised NOS trying to conceal its actions from the application plane of an SDN and does not require a trusted data plane to work. Upon detection of anomalies, Gwardar can apply patches and if required, take over the network. Our performance metrics over simulated settings prove the practicality of our solution. Currently, we aim to extend Gwardar by leveraging the capabilities of stateful SDN data planes and detecting compromised SDN applications based on the data plane extracted trajectories. 

\section*{Acknowledgment}
We acknowledge the useful comments by Distinguished Prof. Yvo Desmedt of University College London and University of Texas at Dallas.

\bibliographystyle{IEEEtran}

\raggedright
\bibliography{ref.bib}

\begin{thebibliography}{10}
\providecommand{\url}[1]{#1}
\csname url@samestyle\endcsname
\providecommand{\newblock}{\relax}
\providecommand{\bibinfo}[2]{#2}
\providecommand{\BIBentrySTDinterwordspacing}{\spaceskip=0pt\relax}
\providecommand{\BIBentryALTinterwordstretchfactor}{4}
\providecommand{\BIBentryALTinterwordspacing}{\spaceskip=\fontdimen2\font plus
\BIBentryALTinterwordstretchfactor\fontdimen3\font minus
  \fontdimen4\font\relax}
\providecommand{\BIBforeignlanguage}[2]{{%
\expandafter\ifx\csname l@#1\endcsname\relax
\typeout{** WARNING: IEEEtran.bst: No hyphenation pattern has been}%
\typeout{** loaded for the language `#1'. Using the pattern for}%
\typeout{** the default language instead.}%
\else
\language=\csname l@#1\endcsname
\fi
#2}}
\providecommand{\BIBdecl}{\relax}
\BIBdecl

\bibitem{nsa2}
``{NSA Preps America for Future Battle, Spiegel},''
  \url{http://www.spiegel.de/international/world/new-snowden-docs-indicate-scope-of-nsa-preparations-for-cyber-battle-a-1013409.html},
  2015.

\bibitem{nsa}
``{Snowden: The NSA planted backdoors in Cisco products, Infoworld},''
  \url{http://infoworld.com/article/2608141/internet-privacy/snowden--the-nsa-planted-backdoors-in-cisco-products.html},
  2014.

\bibitem{shaghaghisurvey}
A.~Shaghaghi, M.~A. Kaafar, R.~Buyya, and S.~Jha, ``Software-defined network
  (sdn) data plane security: Issues, solutions and future directions,''
  \emph{Cluster Computing}, 2018.

\bibitem{ali2015survey}
S.~T. Ali, V.~Sivaraman, A.~Radford, and S.~Jha, ``A survey of securing
  networks using software defined networking,'' \emph{IEEE transactions on
  reliability}, vol.~64, no.~3, pp. 1086--1097, 2015.

\bibitem{dhawan2015sphinx}
M.~Dhawan, R.~Poddar, K.~Mahajan, and V.~Mann, ``Sphinx: Detecting security
  attacks in software-defined networks.'' in \emph{NDSS}, vol.~15, 2015, pp.
  8--11.

\bibitem{kreutz2015software}
D.~Kreutz, F.~M. Ramos, P.~E. Verissimo, C.~E. Rothenberg, S.~Azodolmolky, and
  S.~Uhlig, ``Software-defined networking: A comprehensive survey,''
  \emph{Proceedings of the IEEE}, vol. 103, no.~1, pp. 14--76, 2015.

\bibitem{shin2014rosemary}
S.~Shin, Y.~Song, T.~Lee, S.~Lee, J.~Chung, P.~Porras, V.~Yegneswaran, J.~Noh,
  and B.~B. Kang, ``Rosemary: A robust, secure, and high-performance network
  operating system,'' in \emph{Proceedings of the 2014 ACM SIGSAC conference on
  computer and communications security}.\hskip 1em plus 0.5em minus 0.4em\relax
  ACM, 2014, pp. 78--89.

\bibitem{scott2014operationcheckpoint}
S.~Scott-Hayward, C.~Kane, and S.~Sezer, ``Operationcheckpoint: Sdn application
  control,'' in \emph{Network Protocols (ICNP), 2014 IEEE 22nd International
  Conference on}.\hskip 1em plus 0.5em minus 0.4em\relax IEEE, 2014, pp.
  618--623.

\bibitem{banse2015secure}
C.~Banse and S.~Rangarajan, ``A secure northbound interface for sdn
  applications,'' in \emph{Trustcom/BigDataSE/ISPA, 2015 IEEE}, vol.~1.\hskip
  1em plus 0.5em minus 0.4em\relax IEEE, 2015, pp. 834--839.

\bibitem{rezvani2016anomaly}
M.~Rezvani, A.~Ignjatovic, M.~Pagnucco, and S.~Jha, ``Anomaly-free policy
  composition in software-defined networks.'' in \emph{Networking}, 2016, pp.
  28--36.

\bibitem{jin2015covisor}
X.~Jin, J.~Gossels, J.~Rexford, and D.~Walker, ``Covisor: A compositional
  hypervisor for software-defined networks.'' in \emph{NSDI}, vol.~15, 2015,
  pp. 87--101.

\bibitem{ropke2015sdn}
C.~R{\"o}pke and T.~Holz, ``Sdn rootkits: Subverting network operating systems
  of software-defined networks,'' in \emph{International Workshop on Recent
  Advances in Intrusion Detection}.\hskip 1em plus 0.5em minus 0.4em\relax
  Springer, 2015, pp. 339--356.

\bibitem{tatang2017sdn}
D.~Tatang, F.~Quinkert, J.~Frank, C.~R{\"o}pke, and T.~Holz, ``Sdn-guard:
  Protecting sdn controllers against sdn rootkits,'' in \emph{Network Function
  Virtualization and Software Defined Networks (NFV-SDN), 2017 IEEE Conference
  on}.\hskip 1em plus 0.5em minus 0.4em\relax IEEE, 2017, pp. 297--302.

\bibitem{shaghaghi2017wedgetail}
A.~Shaghaghi, M.~A. Kaafar, and S.~Jha, ``Wedgetail: An intrusion prevention
  system for the data plane of software defined networks,'' in
  \emph{Proceedings of the 2017 ACM on Asia Conference on Computer and
  Communications Security}.\hskip 1em plus 0.5em minus 0.4em\relax ACM, 2017,
  pp. 849--861.

\bibitem{pelekis2010unsupervised}
N.~Pelekis, I.~Kopanakis, C.~Panagiotakis, and Y.~Theodoridis, ``Unsupervised
  trajectory sampling,'' in \emph{Joint European Conference on Machine Learning
  and Knowledge Discovery in Databases}.\hskip 1em plus 0.5em minus 0.4em\relax
  Springer, 2010, pp. 17--33.

\bibitem{orlowski2010sndlib}
S.~Orlowski, R.~Wess{\"a}ly, M.~Pi{\'o}ro, and A.~Tomaszewski, ``Sndlib
  1.0--survivable network design library,'' \emph{Networks}, vol.~55, no.~3,
  pp. 276--286, 2010.

\bibitem{routeview}
{Route Views}, \emph{\url{http://www.routeviews.org}}.

\bibitem{Mausezahn}
``Mausezahn,'' \url{http://www.perihel.at/sec/mz/}.

\bibitem{3}
{CRATE datasets}, \emph{\url{ftp://download.iwlab.foi.se/dataset}}.

\bibitem{4}
{Data Set for IMC 2010 Data Center Measurement},
  \emph{\url{http://pages.cs.wisc.edu/~tbenson/IMC10_Data.html}}.

\bibitem{mizrak2008detecting}
A.~T. M{\i}zrak, S.~Savage, and K.~Marzullo, ``Detecting malicious packet
  losses,'' \emph{IEEE Transactions on Parallel \& Distributed Systems}, no.~2,
  pp. 191--206, 2008.

\end{thebibliography}

\end{document}